\documentclass[twocolumn,aps,prl,groupedaddress,showpacs]{revtex4-1}
\usepackage{amsfonts}
\usepackage{amsmath}
\usepackage{amssymb}
\usepackage{txfonts}
\usepackage{pxfonts}
\usepackage{graphicx,bm,units,yfonts}
\usepackage[table]{xcolor}

\begin{document}

\title{Effect of Strong Disorder on 3-Dimensional Chiral  Topological Insulators: Phase Diagrams, Maps of the Bulk Invariant and Existence of Topological Extended Bulk States}
\author{Juntao Song$^{1,2}$, Carolyn Fine$^{1}$ and Emil Prodan$^{1,\ast}$}
\affiliation{\ $^{1}$Department of Physics, Yeshiva University, New York, NY 10016, USA\\
$^2$Department of Physics, Hebei Normal University, Shijiazhuang, Hebei 050024, China}

\begin{abstract}
The effect of strong disorder on chiral-symmetric 3-dimensional lattice models is investigated via analytical and numerical methods. The phase diagrams of the models are computed using the non-commutative winding number, as functions of disorder strength and model's parameters. The localized/delocalized characteristic of the quantum states is probed with level statistics analysis. Our study re-confirms the accurate quantization of the non-commutative winding number in the presence of strong disorder, and its effectiveness as a numerical tool. Extended bulk states are detected above and below the Fermi level, which are observed to undergo the so called ``levitation and pair annihilation" process when the system is driven through a topological transition. This suggests that the bulk invariant is carried by these extended states, in stark contrast with the 1-dimensional case where the extended states are completely absent and the bulk invariant is carried by the localized states.
\end{abstract}

%\pacs{*****}

\maketitle

\section{Introduction}

The effect of disorder \cite{LiPRL2009xi,GrothPRL2009xi,HJiangPRB2009pd,ProdanPRB2011vy,HMGuoPRL2010pj,JTSongPRB2012pd,ProdanPRB2012dh,YYZhangPRB2012cw,YYZhangPRB2013gw} is well understood for the entire classes A and AII \cite{SchnyderPRB2008qy,KitaevArXiv2009oh,RyuNJP2010tq} of topological insulators ({\it i.e.} all even space dimensions), for both bulk and edge states \cite{BELLISSARD:1994xj,ProdanJPA2013hg,Kellendonk:2002of,Prodan:2009lo}. In the bulk, the existence of a quantized non-trivial topological invariant automatically implies the existence of bulk extended states residing above and below the Fermi energy. Indeed, if such states were absent, then the topological invariants must be zero because, if the entire spectrum is localized, then the Fermi level can be moved all the way to the edges of the spectrum where the topological invariants are de facto zero. Furthermore, the disorder-induced topological-to-trivial transition always happens via the ``levitation" of these extended bulk states \cite{ProdanPRL2010ew,ProdanPRB2012dh,YYZhangPRB2012cw,Prodan:2009lo}  towards each other and through the annihilation of the topological charges carried by these extended states at their collision. An explicit analysis and simulation of these phenomena in 2-dimensional Chern insulators can be found in Ref.~\cite{ProdanJPhysA2011xk}.

For the other unitary class of topological insulators, the chiral or AIII class \cite{SchnyderPRB2008qy,KitaevArXiv2009oh,RyuNJP2010tq}, the equivalent of the non-commutative topological invariant has been recently introduced \cite{MondragonShemArxiv2013ew}: the non-commutative winding number or the odd Chern number. Using similar non-commutative geometry arguments as for the non-commutative Chern number, the non-commutative winding number has been shown \cite{ProdanOddChernArxiv2014} to stay quantized and non-fluctuating (from a disorder configuration to another) even after the spectral gap of the insulator was closed by disorder. Even so, the arguments which led us to the "levitation and pair annihilation" phenomenon in class A break down, because the Fermi level is pinned at $E_F=0$ for chiral-symmetric systems (the definition of the winding number requires that). As such, the existence of extended states in the bulk remains an open problem. The recent studies \cite{MondragonShemArxiv2013ew,SongPRB2014bb} carried in space-dimension $d=1$ have found that the bulk extended states are completely absent even for the topological phases, and that the winding numbers are entirely carried by localized states. Then the important question that emerged is if this situation is generic or if it is specific only to the case $d=1$?

In this paper we consider an explicit 3-dimensional disordered model from the AIII-symmetry class and investigate its phase diagram and the quantum characteristic of the topological states. The goal of our
study is three-fold: 1) We want to demonstrate explicitly the topological properties of the non-commutative winding number in $d=3$ and in the presence of strong disorder, that is, its fine quantization and the non-fluctuating characteristic when the Fermi level is embedded in dense localized spectrum. 2) We want to demonstrate the effectiveness of the non-commutative winding number as a numerical tool. 3) For the topological phases, we want to prove the existence of the bulk extended states (at enormous disorder strengths!) and that the topological transitions proceed through the ``levitation and annihilation" mechanism, as described above.

\section{The 3-Dimensional Model: Definition and Characterization}

\subsection{The Clean Case}

We work with the $Cl_{5,0}$ Clifford algebra:
\begin{equation}
\Gamma_i \Gamma_j + \Gamma_j \Gamma_i =2\delta_{ij}, \ i,j=1,\ldots,5,
\end{equation}
and we choose the following $4 \times 4$ explicit irreducible representation:
\begin{equation}
\begin{array}{c}
 \Gamma_1=\left (^0_{\sigma_1} \ ^{\sigma_1}_0\right),\ \Gamma_2=\left (^0_{\sigma_2} \ ^{\sigma_2}_0\right), \ \Gamma_3=\left (^0_{\sigma_3} \ ^{\sigma_3}_0\right), \medskip \\
\Gamma_4=i\left (^0_{I} \ ^{-I}_0\right), \ \Gamma_5=\left (^I_{0} \ ^{0}_I\right),
\end{array}
\end{equation}
for the $\Gamma$ matrices ($i=1,\ldots,5$). Above, $\sigma_i$'s represent the $2 \times 2$ Pauli's matrices. The models are defined on the space $\ell^2(\mathbb Z^3,\mathbb C^4)$ of square summable functions $\bm \psi_{\bm x}$ defined on the lattice $\mathbb Z^3$ with values in $\mathbb C^4$. The minimal chiral-symmetric topological Hamiltonian that can be built with the aid of these $\Gamma$-matrices takes the following explicit form:
\begin{equation}\label{CleanModelR1}
\begin{array}{c}
(H \bm \psi)_{\bm x} =m\Gamma_4 \bm \psi_{\bm x} \medskip \\
 + \nicefrac{1}{2}\sum_{j=1}^3  \left \{ i\Gamma_j \left (\bm \psi_{\bm x - \bm e_j} - \bm \psi_{\bm x +  \bm e_j} \right )+  \Gamma_4 \left (\bm \psi_{\bm x - \bm e_j} + \bm \psi_{\bm x + \bm e_j} \right ) \right \},
\end{array}
\end{equation}
where $\bm e_j$'s represent the fundamental translations of the lattice. Since $\Gamma_5 \Gamma_j \Gamma_5 = -\Gamma_j$ for $j=1,\ldots,4$, it is evident that $H$ has the chiral-symmetry which is implemented by $\Gamma_5$. Despite its minimality, the model displays a rich phase diagram as a function of the (unique) parameter $m$.

\begin{figure}
\includegraphics[width=8.6cm]{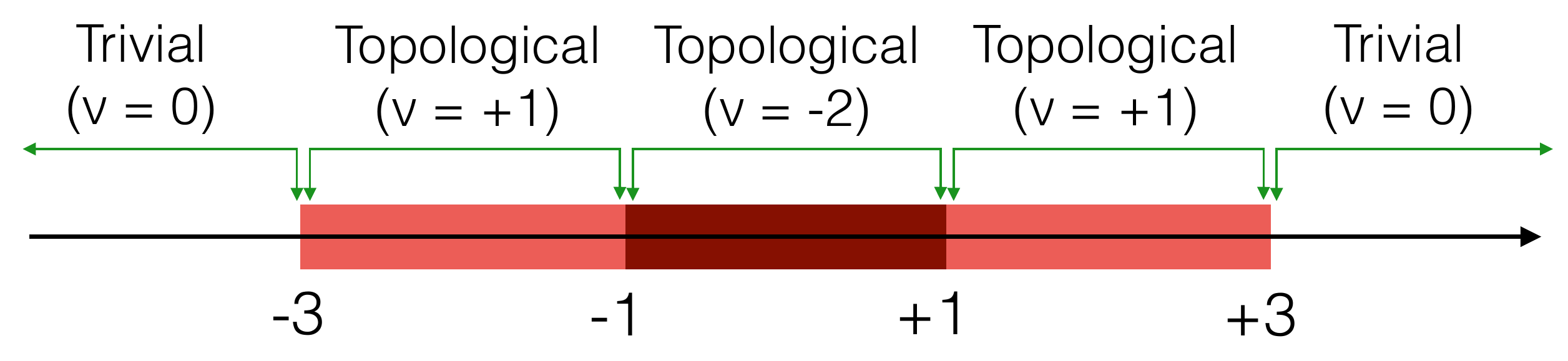}
\caption{(Color online) The phase diagram of the clean model defined in Eq.~\ref{CleanModelR1} or \ref{CleanModelK1}. Here, the reader can identify the topological phases with $\nu=-2$, $\nu=+1$, and the trivial topological phase $\nu=0$ with varying the parameter $m$. }
\label{PhaseDiagramClean1}
\end{figure}

Indeed, in $k$-space, the model takes the explicit form \cite{RyuNJP2010tq}:
\begin{equation}\label{CleanModelK1}
H_{\bm k}=\sum_{j=1}^3 \sin k_j \Gamma_j +(m+\sum_{j=1}^3 \cos k_j ) \Gamma_4.
\end{equation}
Given the defining properties of the $\Gamma$-matrices, one has:
\begin{equation}
H_{\bm k}^2 = \left [ \sum_{j=1}^3 \sin^2 k_j  +(m+\sum_{j=1}^3 \cos k_j )^2 \right ] I_{4 \times 4},
\end{equation}
hence band spectrum:
\begin{equation}
E_{\bm k}^\pm = \pm \left [ \sum_{j=1}^3 \sin ^2 k_j +(m+\sum_{j=1}^3 \cos k_j )^2 \right ]^\frac{1}{2}
\end{equation}
and the flat-band Hamiltonian $Q_{\bm k} \equiv \frac{H(\bm k)}{|H(\bm k)|}$:
\begin{equation}
Q_{\bm k} =(E_{\bm k}^+)^{-1}\left[\sum_{j=1}^3 \sin k_j \Gamma_j +(m+\sum_{j=1}^3 \cos k_j ) \Gamma_4 \right ],
\end{equation}
can be explicitly computed. This $Q_{\bm k}$ has only off-diagonal terms (due to the chiral symmetry): $Q_{\bm k}=\left( ^0_{U_{\bm k}^\dagger} \ ^{U_{\bm k}}_0 \right)$, and the unitary matrix $U_{\bm k}$, which uniquely determines the ground state of the model, can be easily read from here:
\begin{equation}
U_{\bm k}=(E_{\bm k}^+)^{-1}\left[\sum_{j=1}^3 \sin k_j \sigma_j -i (m+\sum_{j=1}^3 \cos k_j )I_{2 \times 2}  \right ].
\end{equation}
The bulk invariant is given by the winding number of $U_{\bm k}$ \cite{Schnyder:2009qa}:
\begin{equation}\label{WindingK}
\nu (U_k) = \Lambda_3 \sum_{\rho \in S_3} (-1)^\rho \int_{BZ} d^3 \bm k \prod_{j=1}^3 U_{\bm k}^\dagger \partial_{\rho_j} U_{\bm k},
\end{equation}
where the summation is over all permutations of the three indices. A map of $\nu$ as a function of the parameter $m$ is reported in Fig.~\ref{PhaseDiagramClean1}. As one can see, there are three domains of topological phases with $\nu=+1$ and $\nu = -2$, the transition points being located at $m=-3$, $-1$, $+1$, $3$. At these points, the spectral gap of the model closes.

\begin{figure}
\includegraphics[width=8.6cm]{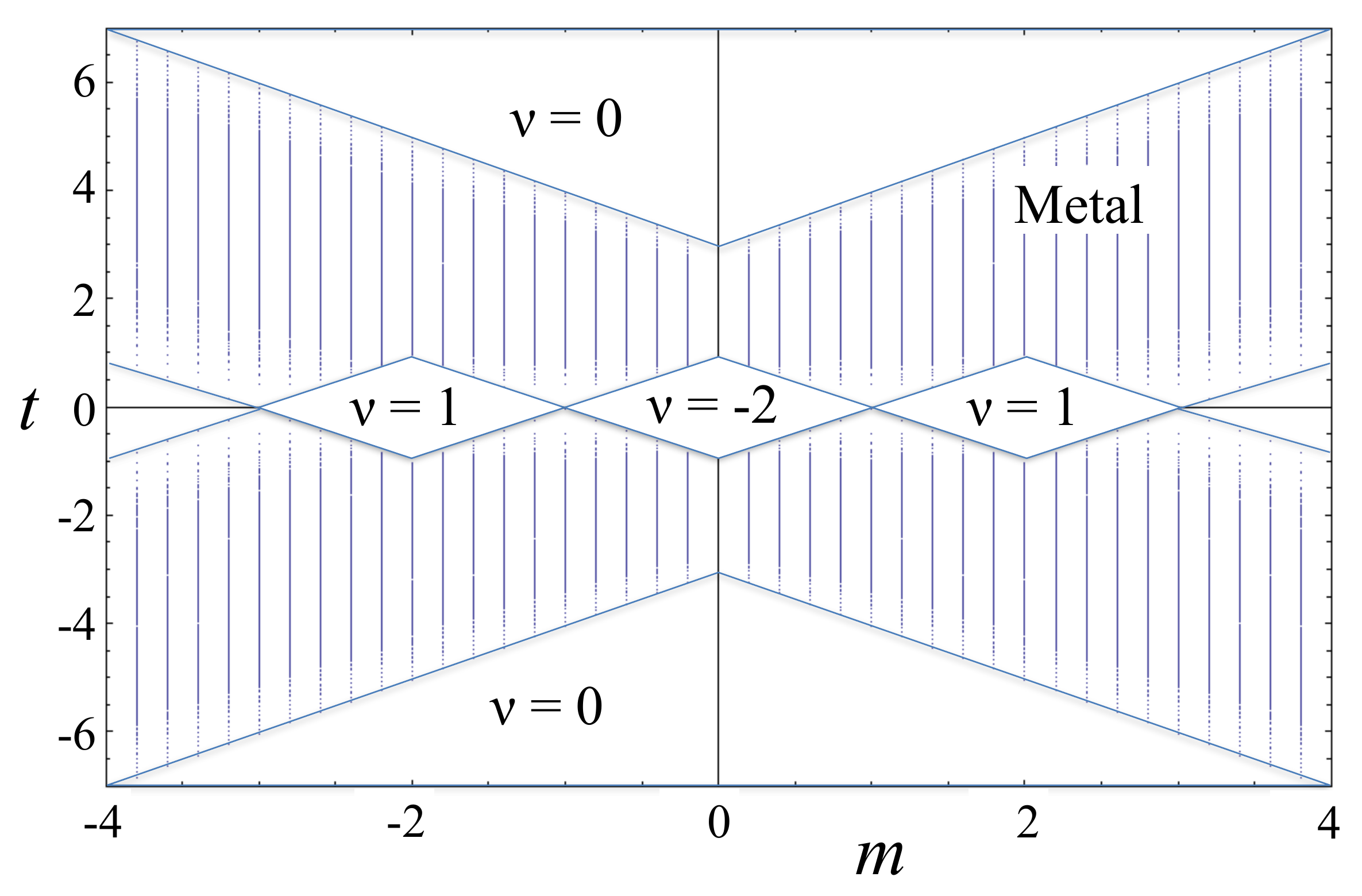}
\caption{(Color online) The phase diagram of the clean model defined in Eq. \ref{CleanModelR2}. Here, the reader can identify the topological phases with $\nu=-2$ and $\nu=+1$ (the rhombic domains), a large metallic phase (the shaded region) and the trivial topological phase $\nu=0$. }
\label{PhaseDiagramClean2}
\end{figure}

The minimal model of Eq.~\ref{CleanModelR1} has two more symmetries: the time-reversal symmetry implemented by $ (\sigma_1 \otimes i \sigma_2) \mathcal K$ (squaring to $-1$) and the particle hole symmetry implemented by $ (\sigma_2 \otimes \sigma_2) \mathcal K$, where $\mathcal K$ is the ordinary complex conjugation operator. Note that these two symmetrical operators do not commute with each other henceforth the model cannot be placed in the DIII-symmetry class \cite{SchnyderPRB2008qy,KitaevArXiv2009oh,RyuNJP2010tq}. It is quite interesting to investigate what happens if we break these symmetries. As such we add one more term to the model, which becomes:
\begin{equation}\label{CleanModelR2}
\begin{array}{c}
 (H_0 \bm \psi)_{\bm x} =m\Gamma_4 \bm \psi_{\bm x} + i t \Gamma_1 \Gamma_3 \Gamma_4 \bm \psi_{\bm x}\medskip \\
 + \nicefrac{1}{2}\sum_{j=1}^3  \left \{ i\Gamma_j \left (\bm \psi_{\bm x - \bm e_j} - \bm \psi_{\bm x +  \bm e_j} \right )+  \Gamma_4 \left (\bm \psi_{\bm x - \bm e_j} + \bm \psi_{\bm x + \bm e_j} \right ) \right \}.
\end{array}
\end{equation}
In $k$-space, the extended model takes the form:
\begin{equation}\label{CleanModelK2}
H_{\bm k}=\sum_{j=1}^3 \sin k_j \Gamma_j +(m+\sum_{j=1}^3 \cos k_j ) \Gamma_4 +i t \Gamma_1 \Gamma_3 \Gamma_4,
\end{equation}
and the bulk invariant can be computed as before. A map of the winding number for the model in Eq.~\ref{CleanModelR2} is reported in Fig.~\ref{PhaseDiagramClean2}. The most important feature in this diagram is the emergence of a metallic (gapless) phase which now surrounds the domains of topological phases.

The last comment for this section is that both models are interesting for our analysis in the presence of disorder. Indeed, while the time-reversal and particle-hole symmetries do not play any topological role, as we shall see, their presence or absence moves the critical points between the topological phases from the symplectic universal class to the unitary universal class, which can induce distinct physically measurable effects.

\begin{figure}
\includegraphics[width=7.0cm]{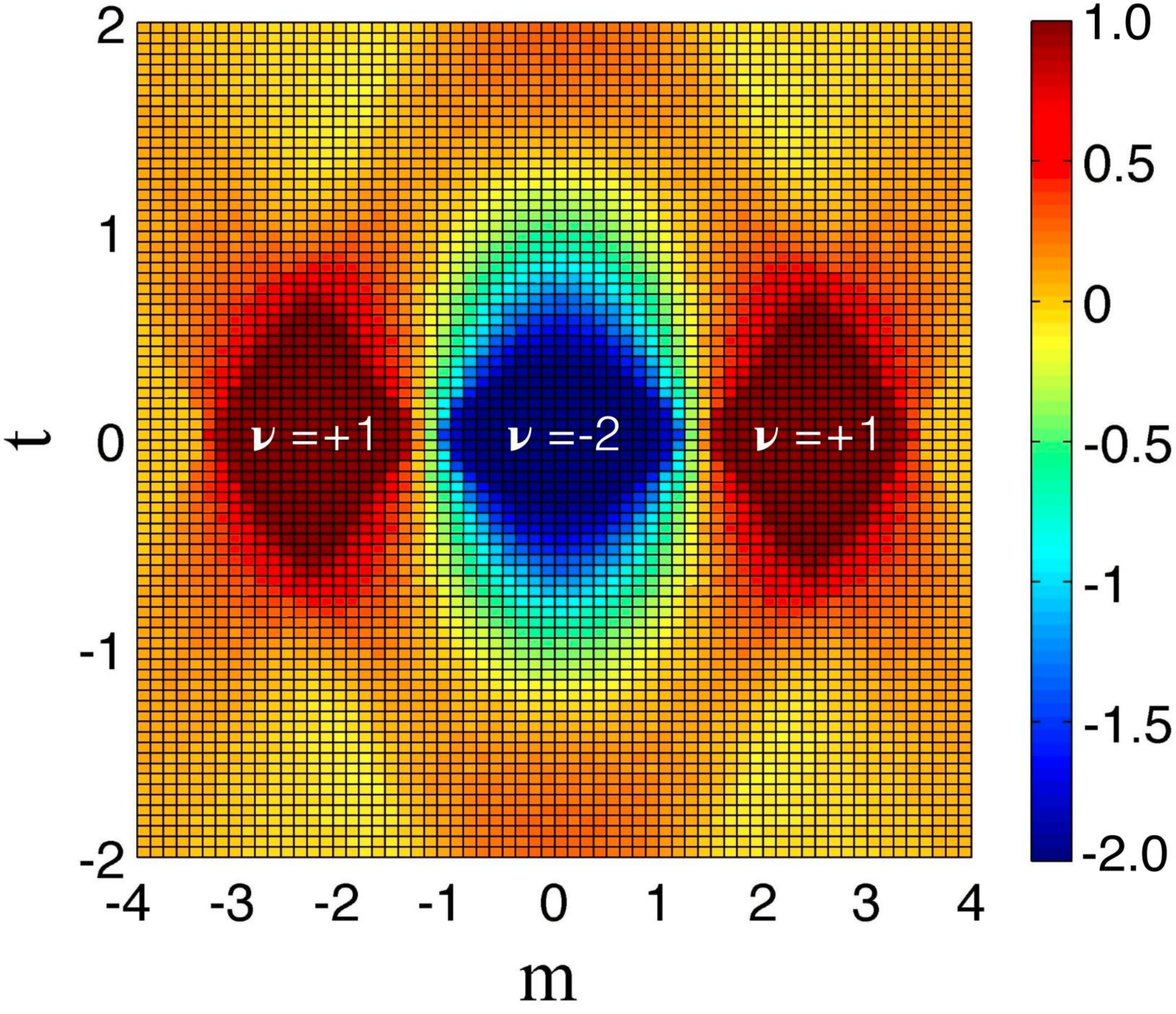}
\caption{(Color online) The phase diagrams in the phase space $(m,t)$ at disorder strength $W=4$. The computations were completed on a cubic lattice of $N=16\times 16 \times 16$ unit cells, following the procedure described in the text.}\label{PhaseDiagramDisorder1}
\end{figure}

\subsection{The Disordered Case}

We only consider on-site disorder, induced by random fluctuations of $m$:
\begin{equation}\label{DisorderedModel1}
\begin{array}{c}
 (H_\omega \bm \psi)_{\bm x} =(m+W \omega_{\bm x}) \Gamma_4 \bm \psi_{\bm x}  + i t \Gamma_1 \Gamma_3 \Gamma_4 \bm \psi_{\bm x}\medskip \\
+\nicefrac{1}{2}\sum_{j=1}^3  \left \{ i \Gamma_j \left (\bm \psi_{\bm x - \bm e_j} - \psi_{\bm x +  \bm e_j} \right )+  \Gamma_4 \left (\bm \psi_{\bm x + \bm e_j} + \bm \psi_{\bm x - \bm e_j} \right ) \right \},
\end{array}
\end{equation}
where $\{\omega_{\bm x}\}_{x\in \mathbb Z^3}$ are independent random numbers drawn from the interval $[-\nicefrac{1}{2},\nicefrac{1}{2}]$ (white noise). As one can easily see, the disordered Hamiltonian continues to display the chiral-symmetry: $\Gamma_5 H_\omega \Gamma_5 = -H_\omega$.

The following details are of technical nature but nevertheless important for our analysis, and the related mathematic argumentation can be also found in Refs. ~\cite{SongPRB2014bb,ProdanAMRX2013bn,ProdanOddChernArxiv2014}. Readers who are only interested in physical results on the first reading can skip the following mathematic part.  We denote a generic disorder configuration $\{\omega_{\bm x}\}_{x\in \mathbb Z^3}$ by $\omega$, and the latter is seen as a point in $\Omega = [-\nicefrac{1}{2},\nicefrac{1}{2}]^{\mathbb Z^3}$. This is compact metrizable set which admits a probability measure, to be used for disorder averages, which is simply defined by $dP(\omega)=\prod_{\bm x \in \mathbb Z^3} d \omega_{\bm x}$. It is important to note that there is a natural action of the lattice translations on $\Omega$:
\begin{equation}
(t_{\bm a} \omega)_{\bm x} = \omega_{\bm x + \bm a},
\end{equation}
and that the measure $dP(\omega)$ is ergodic relative to this action. The family of disordered Hamiltonians $\{H_\omega\}_{\omega \in \Omega}$ defined in Eq.~\ref{DisorderedModel1} is covariant, in the sense that:
\begin{equation}
T_{\bm a}H_\omega T_{\bm a}^{-1} = H_{t_{\bm a}\omega},
\end{equation}
for any lattice translation $T_{\bm a}$. Furthermore, any family of operators $\{\phi(H_\omega)\}_{\omega \in \Omega}$ produced by the functional calculus with $H_\omega$ is covariant, and the same can be said for the commutators $\{[\bm X, \phi(H_\omega)]\}_{\omega \in \Omega}$, where $\bm X$ is the position operator. The covariant property, together with the ergodicity of the probability measure, ensures the following self-averaging principle:
\begin{equation}\label{Birkhoff}
\mathcal T\{F_\omega G_\omega \ldots\} = \int_\Omega d\omega \ \mathrm{tr}_{\bm 0} \{F_\omega G_\omega \ldots\}
\end{equation}
for any covariant observables $F_\omega, G_\omega, \ldots$. Above, $\mathcal T\{\cdot \}$ represents the trace per volume and $\mathrm{tr}_0$ is the trace over $\mathbb C^4$.

\begin{figure}
\includegraphics[width=7.0cm]{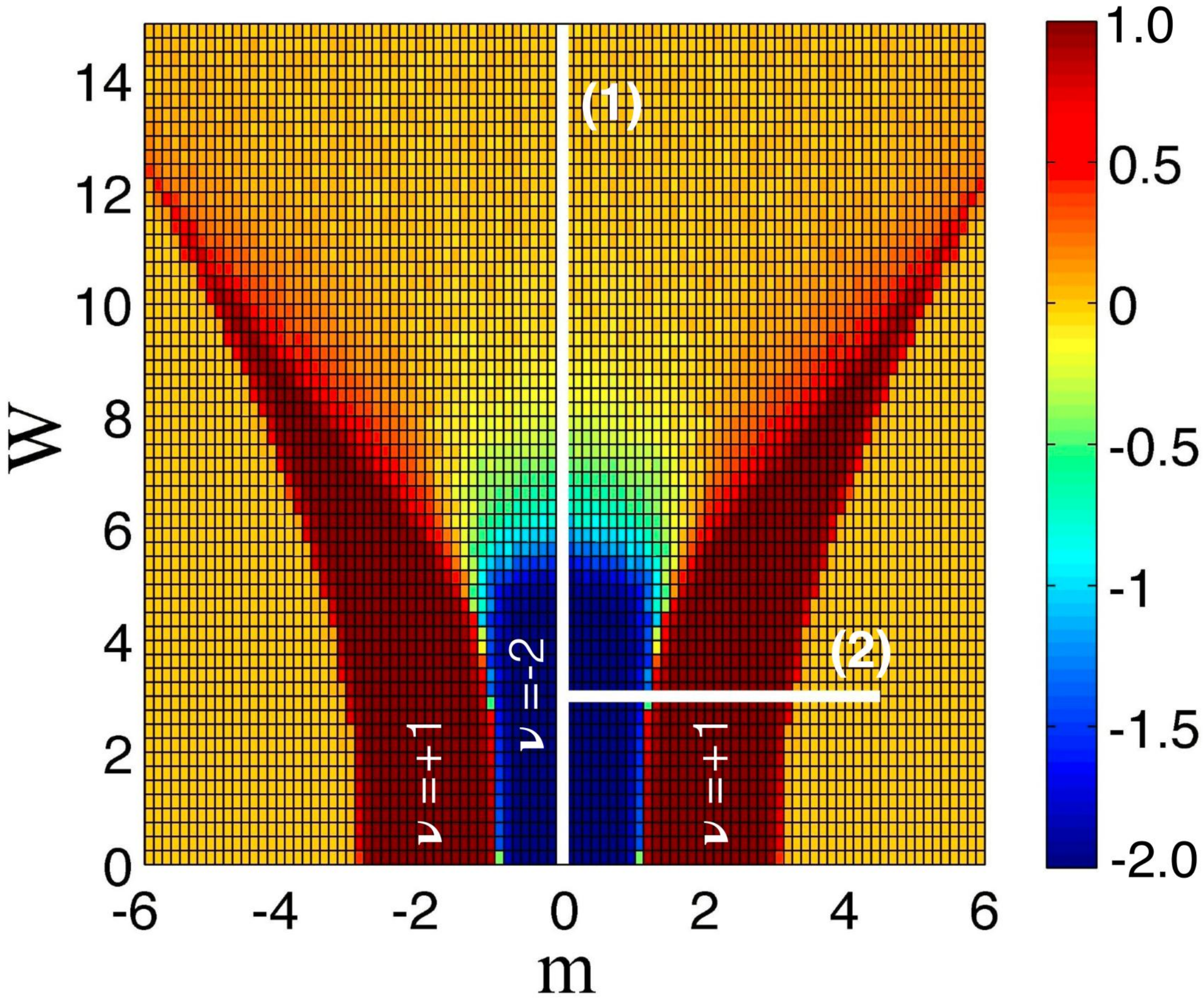}
\caption{(Color online) The phase diagrams in the phase space $(m,W)$ at $t=0$. The computations for $\nu$ were done with a cubic lattice of $N=16\times 16 \times 16$ unit cells.}\label{PhaseDiagramDisorder2}
\end{figure}

The bulk invariant can be defined as before, with the only difference that the calculus must proceed in the real space representation. Indeed, by considering again the flat-band Hamiltonian, the chiral symmetry annihilates the diagonal blocks and:
\begin{equation}
Q_\omega = \frac{ H_\omega}{|H_\omega|}= \left (
\begin{array}{cc}
0 & U_\omega \\
U_\omega^\dagger & 0
\end{array}
\right )
\end{equation}
with $U_\omega$ a unitary operator which generates a covariant family when $\omega$ is allowed to take values in $\Omega$. The natural generalization of the winding number to the disordered case is:
\begin{equation}\label{WindingR}
\nu (U_\omega) = \frac{i \pi}{3} \sum_{\rho \in S_3} (-1)^\rho \ \mathcal T  \left \{\prod_{j=1}^3 U_\omega^{-1} [X_{\rho_j},U_\omega]  \right \},
\end{equation}
which for the translational invariant case is just the real space representation of the $k$-space formula in Eq.~\ref{WindingK}. The following index theorem is adopted from Ref.~\cite{ProdanOddChernArxiv2014}: \smallskip

\noindent {\bf Theorem \cite{ProdanOddChernArxiv2014}} On the space $\ell^2(\mathbb Z^3,\mathbb C^4)\otimes \mathbb C^2$ let $\sum_{j=1}^3 X_j \otimes \sigma_j$ be the Dirac operator and let $\Pi$ denote the projector onto the positive spectrum of this Dirac operator. Assume:
\begin{equation}\label{cond}
\int_\Omega d P(\omega) \ |\langle \bm x|U_\omega|\bm y\rangle| \leq Ae^{-\gamma |\bm x - \bm y|},
\end{equation}
for some strictly positive $A$ and $\gamma$. Then, with probability one in $\omega$, $\Pi U_\omega \Pi$ is a Fredholm operator and:
\begin{equation}
\nu (U_\omega) = \mathrm{Index} \ \Pi U_\omega \Pi.
\end{equation}
Furthermore, the Fredholm index on the righthand side is independent of $\omega$ and is invariant against any continuous deformations of the Hamiltonian as long as Eq.~\ref{cond} is satisfied.\smallskip

\begin{figure}
\includegraphics[width=8.6cm]{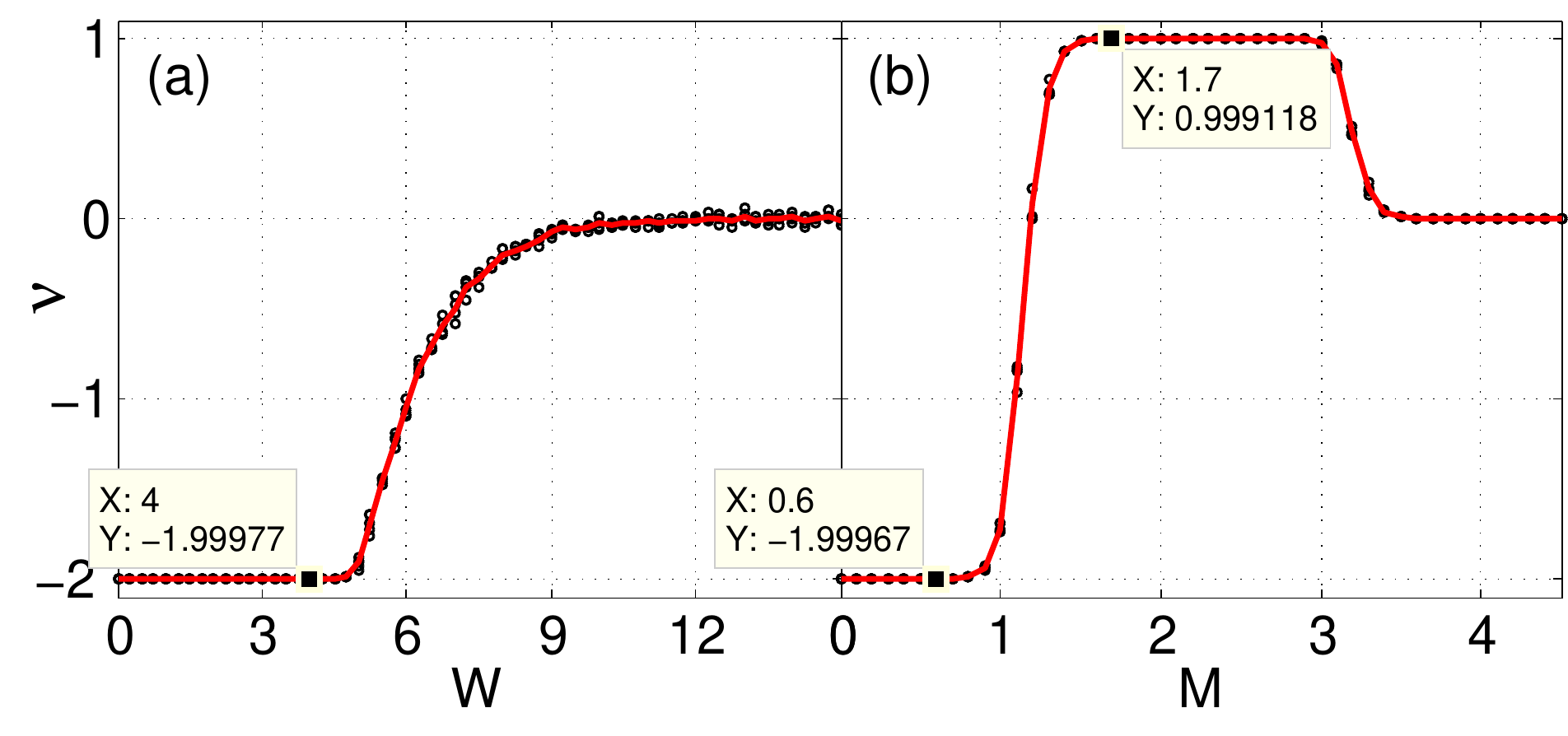}
\caption{(Color online) Evolution of the winding number $\nu$ with disorder W (a) and parameter m (b). The raw, un-averaged data for 5 disorder configurations is shown by the scattered points and the average by the solid line. The marked data points report quantized values with 3 digital precisions.}\label{WindingNumber}
\end{figure}

\begin{figure*}
\includegraphics[width=12cm]{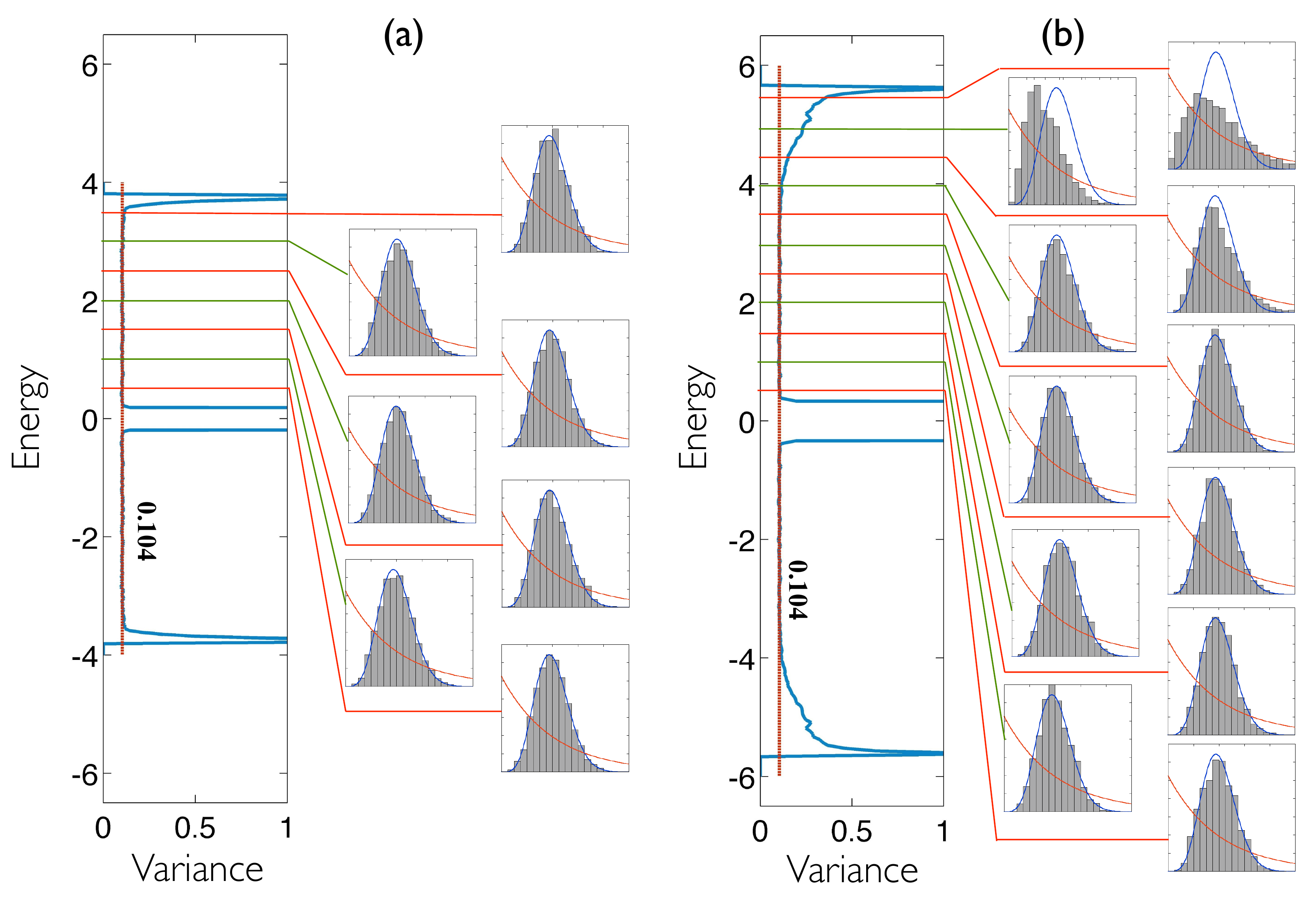}
\caption{(Color online) Statistics of the energy level-spacing ensembles for Hamiltonian defined in Eq.~\ref{DisorderedModel1} with $t=0$ (time-reversal symmetry) and disorder strength $W=4$, collected at different energies. Section (a) of the figure corresponds to the topological phase $\nu=-2$ ($m=0$), while section to the topological phase $\nu=+1$ ($m=2$). For both sections, the main panels show the variance of the ensembles. The dotted lines mark the value $0.104$ appropriate for a Gaussian symplectic ensemble of random matrices. The side panels show the histograms of the level-spacings ensembles recorder at a few particular energies. This histograms are compared with the Wigner surmise distribution $P_{\mbox{\tiny{GSE}}}(s)=\frac{2^{18}}{3^6\pi^3}s^4 e^{-\frac{64}{9\pi}s^2}$.}\label{TRSLevelStatistics}
\end{figure*}

The condition written in Eq.~\ref{cond} holds true if the Fermi level resides in a region of Anderson localized energy spectrum \cite{ProdanOddChernArxiv2014}. This analytic result ensures that the topological phases do not disappear when the disorder is turned on, and that topological phases with different $\nu$'s are separated by a metallic phase boundary. The numerical algorithm we use to compute the non-commutative winding number is based on the canonical finite-volume approximations discovered in Ref.~\cite{ProdanAMRX2013bn} and was discussed in detailed in Ref.~\cite{SongPRB2014bb}. Note that the winding number formula in Eq.~\ref{WindingR} has the self-averaging property discussed above, hence the quantized values of $\nu$ can be obtained from a single disorder configuration, provided the size of the system is large enough. This will prove to be a great numerical advantage of the approach.

Fig.~\ref{PhaseDiagramDisorder1} reports the map of the winding number in the $(m,t)$ plane, computed at fixed disorder strength $W=4$. As one can clearly see, there are well defined regions where the winding number remains quantized and the topological phases seen in Fig.~\ref{PhaseDiagramClean2} are still clearly visible. The phase boundaries of the phase diagram moved quite visibly when compared with Fig.~\ref{PhaseDiagramClean2}, with the topological phases actually occupying more volume after the disorder was turned on. Outside the topological regions, the winding number does not drop to zero immediately, indicating the presence of a substantial metallic region (defined as having a diverging dynamical localization length). Hence, the metallic phase present in  Fig.~\ref{PhaseDiagramClean2} survives the disorder, but this is of course not a surprise in space dimension $d=3$.

Fig.~\ref{PhaseDiagramDisorder2} reports the map of the winding number in the plane $(m,W)$, computed at $t=0$. As one can see, the phase boundaries are strongly affected by the disorder. The topological phases survive up to the extreme disorder strengths of $W= 13$ for $\nu=1$ and $W=7$ for $\nu=-2$. We want to point out the markedly different topology of the phase diagram reported in Fig.~\ref{PhaseDiagramDisorder2} when compared with the phase diagram of the 1-dimensional chiral model reported in Ref.~\cite{SongPRB2014bb} (see Fig.~6a). Although both diagrams look similar at $W=0$, for the 1-dimensional model the $\nu=+1$ phase fully surrounds the other topological phase (which in that case is $\nu=+2$), while in the present case, the $\nu=+1$ phase is actually repelled by the other topological phase $\nu=-2$.

Together, Figs.~\ref{PhaseDiagramDisorder1} and \ref{PhaseDiagramDisorder2} provide a good guidance on how the 3-dimensional phase diagram in the phase space $(m,t,W)$ might look. From such exercise, it is easy to see that the topological phases in this 3-dimensional phase diagram are surrounded by a true metallic phase, which can explain the slow (i.e. not sharp) decay of $\nu$ to zero once it exists the topological phases at large $W$'s (clearly visible in Fig.~\ref{PhaseDiagramDisorder2}). Besides, it should be explicitly pointed out that the boundaries of the topological phase are greatly deformed by disorder, which can be clearly seen in Fig.~\ref{PhaseDiagramDisorder2} (also Fig.~\ref{PhaseDiagramDisorder1}). This deformation of topological phase boundaries enable us observe a intriguing topological phase transition from a trivial phase to non trivial one with increasing disorder strength, e.g. $m=4$ in \ref{PhaseDiagramDisorder2}. This fascinating phenomenon was intensively studied before and confirmed to exist in 1-dimensional  \cite{MondragonShemArxiv2013ew,SongPRB2014bb,Altland1DChiralPRL2014}, 2-dimensional \cite{LiPRL2009xi,GrothPRL2009xi,HJiangPRB2009pd,ProdanPRB2011vy,JTSongPRB2012pd,ProdanPRB2012dh,YYZhangPRB2012cw,YYZhangPRB2013gw}, and 3-dimensional \cite{HMGuoPRL2010pj,Sbierski3DTIPRB2014} cases. More often, people would like to call this topological phase induced by disorder as topological Anderson insulator.

In order to illustrate the quality of the data that can be obtained with the non-commutative winding number, in Fig.~\ref{WindingNumber} we report the numerical values of the winding number along the paths (1) and (2) shown in Fig.~\ref{PhaseDiagramDisorder2}. In this figure we show the results for five independent random configurations (the markers) as well as the average over these five random configurations. The calculations have been completed on a larger lattice of size $21 \times 21 \times 21$. Several explicit numerical values of the averaged winding numbers are displayed, showing a quantization with 3 digits of precision. The data also show the self-averaging property of the winding number, which can be deduced from the absence of fluctuations in the non-averaged data.

\section{The localized/delocalized characteristic of the quantum states}

The localized/delocalized characteristic of the quantum states can be probed by examining the statistics of the energy level spacings \cite{EfetovBook1997vn}. Since the only required inputs are the eigenvalues of the disordered Hamiltonians, this technique is fairly efficient and well suited for mapping large phase diagrams. We closely follow the prescription reported in Ref.~\cite{ProdanJPhysA2011xk}, which consists of recording energy level spacings from small windows centered at different energies (as oppose to treating the whole spectrum at once). This is especially useful when the localized/delocalized characteristic of the quantum states changes with the energy. In the following numerical experiments, we used a random number generator to build the $H_\omega$'s on a $21 \times 21 \times 21$ lattice with periodic boundary conditions. The eigenvalues were collected for 100 disordered configurations and for each energy, the width of the energy window was adjusted so that at the end 2000 level spacings were recorded for each energy. The histograms of these ensembles of level spacings were constructed and the variance of these histograms was computed, following the same method as that in Ref.~\cite{ProdanJPhysA2011xk}.

\begin{figure*}
\includegraphics[width=12cm]{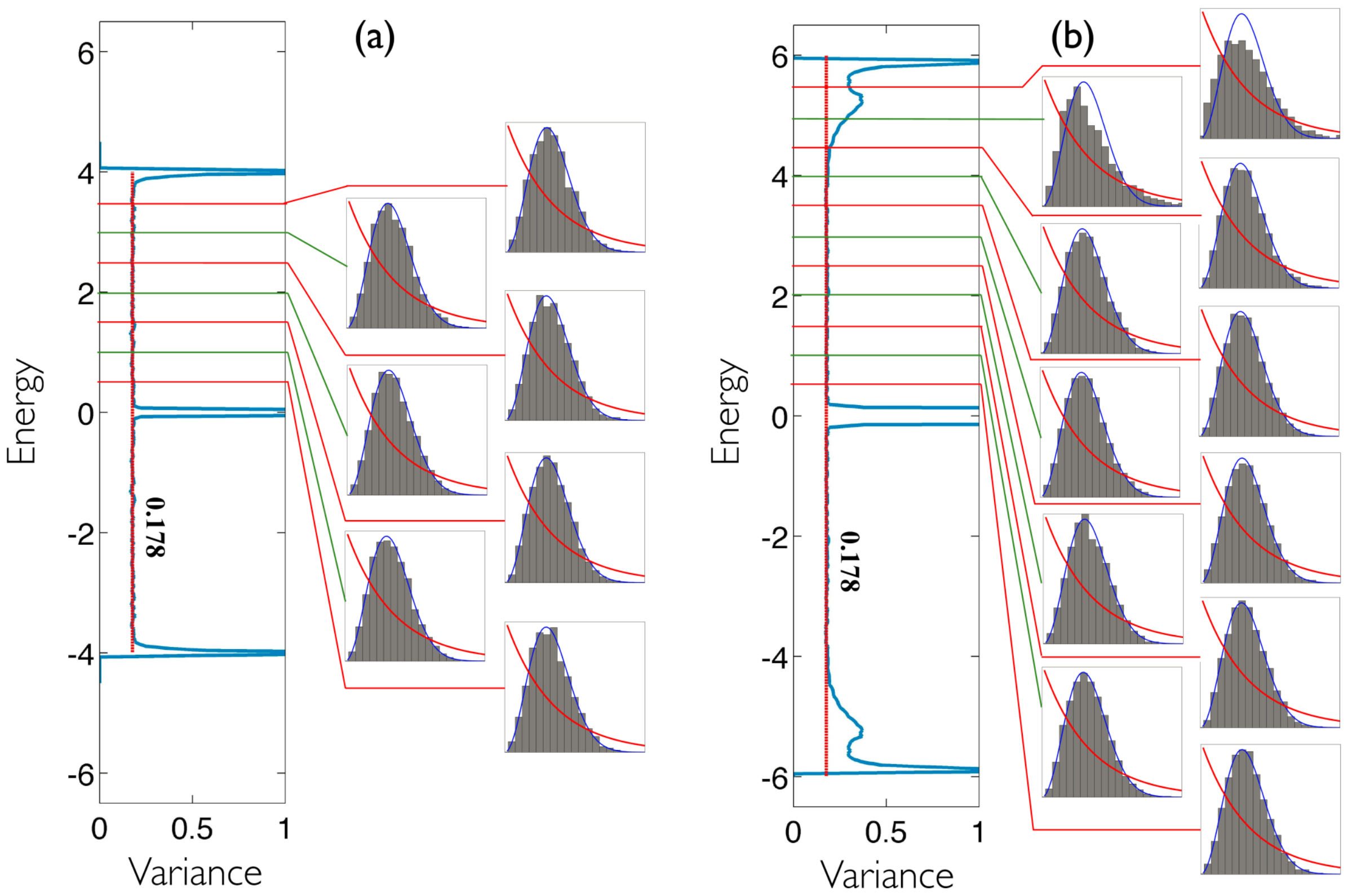}
\caption{(Color online) Statistics of the energy level-spacing ensembles for Hamiltonian defined in Eq.~\ref{DisorderedModel1} with $t=0.3$ (broken time-reversal symmetry) and disorder strength $W=4$, collected at different energies. Section (a) of the figure corresponds to the topological phase $\nu=-2$ ($m=0$), while section to the topological phase $\nu=+1$ ($m=2$). For both sections, the main panels show the variance of the ensembles. The dotted lines mark the value $0.178$ appropriate for a Gaussian unitary ensemble of random matrices. The side panels show the histograms of the level-spacing ensembles recorded at a few particular energies. This histograms are compared with the Wigner surmise distribution $P_{\mbox{\tiny{GUE}}}(s)=\frac{32}{\pi^2}s^2 e^{-\frac{4}{\pi}s^2}$.}\label{BTRSLevelStatistics}
\end{figure*}

\subsection{Existence of extended states}

Here we comb the entire energy spectrum of the models using the level spacing statistics, in search for the bulk extended states above and below the Fermi level $E_F=0$.

Fig.~\ref{TRSLevelStatistics} refers to the Hamiltonian defined in Eq.~\ref{DisorderedModel1} with $t=0$, in which case the time-reversal symmetry is restored. The disorder strength was fixed at $W=4$. Section (a) of this figure refers to the topological phase $\nu=-2$ ($m=0$) and section (b) to the topological phase $\nu=+1$ ($m=2$). For both sections, the main panels report the variance of the energy level ensembles recorded at various energies. As one can see, for the most part of the energy range, the variance is pinned to the value $0.104$, which is the expected value for a Gaussian symplectic ensemble. At the edges of the spectrum the variance approaches the value 1, appropriate for a Poisson distribution (this is more visible in section (b) of the figure). When the variance is pinned at $0.104$, the histograms of the ensembles, shown in the side panels, overlap almost perfectly with the Wigner surmise distribution $P_{\mbox{\tiny{GSE}}}(s)=\frac{2^{18}}{3^6\pi^3}s^4 e^{-\frac{64}{9\pi}s^2}$. According to Ref.~\cite{EfetovBook1997vn}, this is an indication that the dynamical localization length of the states exceeds the simulation box. To obtain the energy region where the dynamical localization length is infinite one needs to perform a finite-size scaling analysis and detect the critical point (if any) which separates the localized and the extended spectrum. While we did not carried this analysis entirely, we did look at the variance for different lattice sizes (the $21 \times 21 \times 21$ lattice was the largest we considered) and observed that the domain where the variance is pinned at $0.104$ is practically not affected by the size. This assured us that here we are in fact dealing with true extended states.

Fig.~\ref{BTRSLevelStatistics} refers to the case when $t=0.3$, in which case both the time-reversal symmetry and the particle hole symmetry are broken. As before, section (a) refers to the topological phase $\nu=-2$ ($m=0$) and section (b) to the topological phase $\nu=+1$ ($m=2$). For this case, one can see the variance being pinned at $0.178$, which is the expected value for a Gaussian unitary ensemble. The histograms of the level spacing ensemble confirm that indeed the distributions follow the Wigner surmise distribution  $P_{\mbox{\tiny{GUE}}}(s)=\frac{32}{\pi^2}s^2 e^{-\frac{4}{\pi}s^2}$. As before, this leaves little doubt that we are again dealing with extended states.

\subsection{Levitation and pair annihilation at the topological transition}

\begin{figure}
\includegraphics[width=8cm]{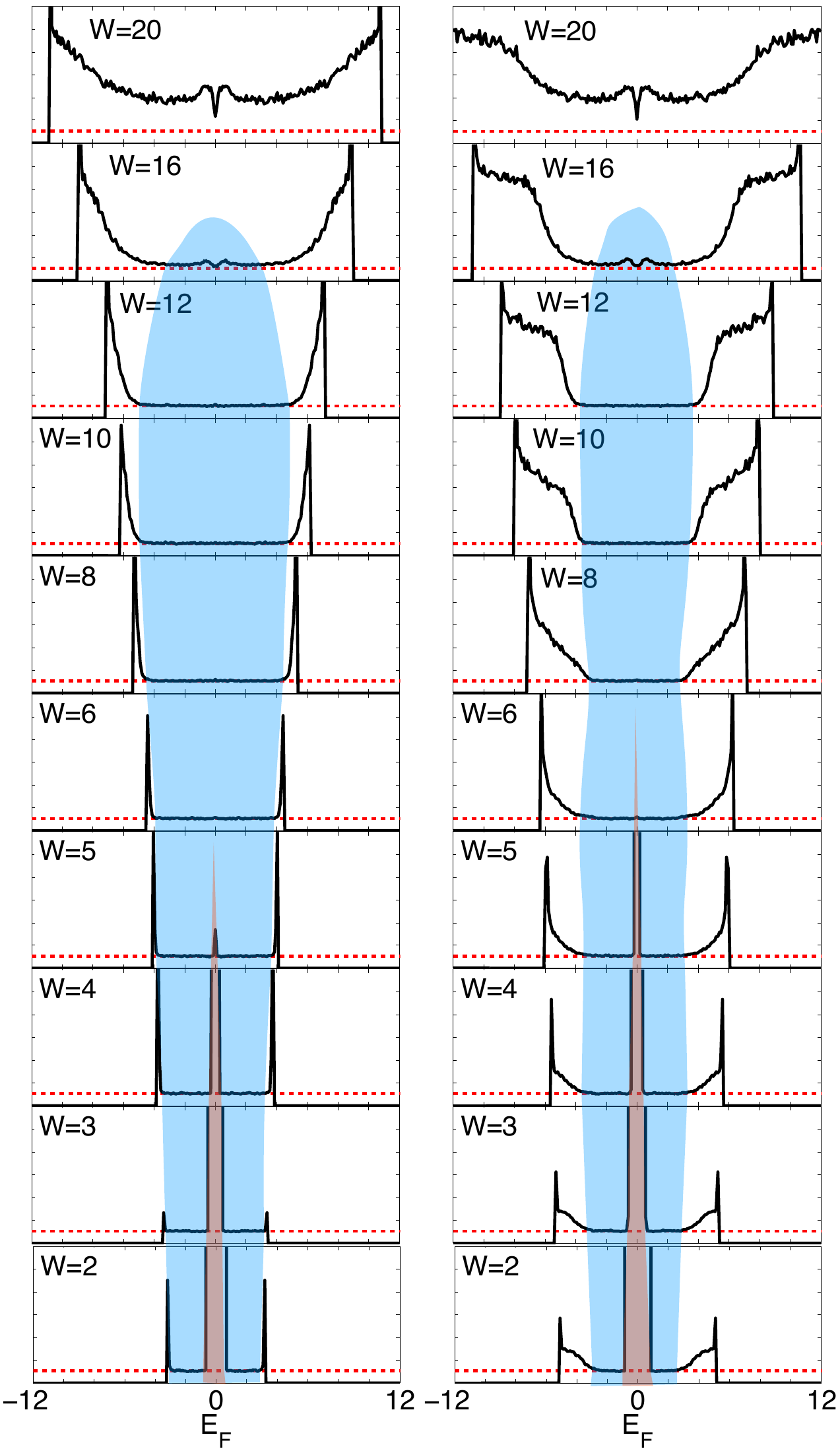}
\caption{(Color online) The variance of the ensembles of level
spacings for the topological case m=0 (left panel) and 2 (right panel), recorded at various disorder strength as a function of Fermi energy. The dotted lines mark the value $0.104$ appropriate for a Gaussian symplectic ensemble of random matrices. A total of 100 disorder configurations were used in these simulations and, for each disorder configuration and energy E, 20 level spacings were collected from the immediate vicinity of E. As such, the ensembles contain 2000 level spacings. The size of the lattice for these simulations was $16\times 16\times 16$.}\label{LevelStatistics}
\end{figure}

The previous data give strong evidence that, indeed, extended bulk states are present above and below the Fermi level. Since the models are in space dimension $d=3$, this in itself is not that surprising. However, when varying the disorder strength and forcing the systems to go through a topological transition, we observed the classical signature of the ``levitation and pair annihilation" phenomenon. Indeed, in Fig.~\ref{LevelStatistics} we report the evolution of the variance of level spacing ensembles as the disorder strength is increased from $W=0$ to $W=20$. As one can clearly see,  the energy domains above and below $E_F=0$ where the variance is pinned at $0.104$ (see the blue-shaded regions in Fig.~\ref{LevelStatistics}) do not disappear as the disordered strength is increased, but instead they move towards each other until they collide and only after the collision they disappear. We point out that the phase diagram presented in Fig.~\ref{LevelStatistics} looks very similar with the phase diagram of the 3-dimensional strong topological insulator investigated in Ref.~\cite{LeungPRB2012vb} (see Fig.~3 there).

When the time-reversal symmetry and the particle hole symmetry are broken, which leads to the degeneracy of eigenvalues, the variance of the energy level ensembles turns to a Gaussian unitary statistics. As shown in Fig.~{LevelStatistics2}, the variance is pinned again to the value $0.178$ for $t=0.3$, which corresponds to extended states for a Gaussian unitary ensemble. It should be pointed out explicitly that with breaking the time-reversal symmetry and the particle hole symmetry the ``levitation and pair annihilation" phenomenon of extended states is still observed with increasing disorder strength, similar as that in Fig.~\ref{LevelStatistics}. Therefore, no matter whether the time-reversal and particle hole symmetries persist or not, the bulk extended states for 3-dimensional chiral topological insulators definitely undergo the ``levitation and pair annihilation" when the system is driven through a topological phase transition.

\begin{figure}
\includegraphics[width=8cm]{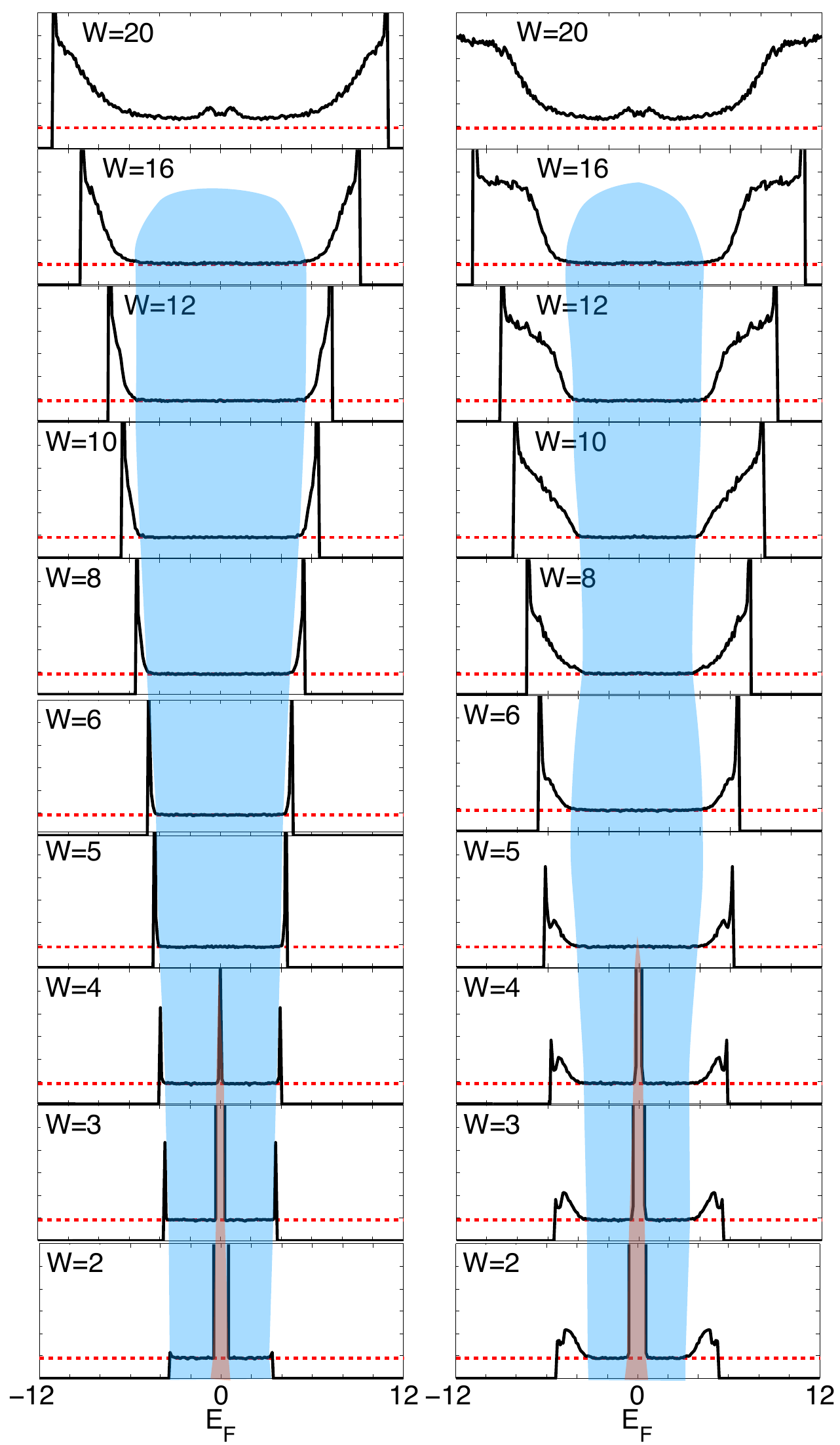}
\caption{(Color online) Same as Fig.~\ref{LevelStatistics}, except for $t=0.3$ (broken time-reversal symmetry). The dotted lines mark the value $0.178$ appropriate for a Gaussian unitary ensemble of random matrices.}\label{LevelStatistics2}
\end{figure}

\section{Conclusions}
Using analytical and numerical methods, we studied the effect of strong disorder on 3-dimensional chiral topological insulators, which follows a $\mathbb Z$-classification.
The main conclusions includes as follows:
\begin{itemize}
\item The non-commutative winding number continues to be an accurate and effective numerical tool in space dimension $d=3$.
\item The bulk extended states survive the disorder even at extreme disorder strengths.
\item The bulk extended states undergo the ``levitation and pair annihilation" when the system is driven through a topological phase transition.
\item This provide strong evidence that the bulk topological invariant is carried by these extended bulk states.
\end{itemize}
These results are helpful to understand topological phase transitions and strong disorder effects for the 3-dimensional topological insulators.

\section{Acknowledgments}
The authors acknowledge extremely fruitful discussion with Ian Mondragon, Taylor Hughes and Hermann Schulz-Baldes. This work was supported by the U.S. NSF grants DMS-1066045, DMR-1056168, NSFC under grants No. 11204065 and NSF-Hebei Province under grants No. A2013205168.

\end{document}